\documentclass[conference,a4paper]{APSIPA2012}
\usepackage{multirow}
\usepackage[dvips]{graphicx}
\usepackage{amsmath}
\usepackage[psamsfonts]{amssymb}
\usepackage{amsxtra}
\usepackage{threeparttable}
\usepackage{graphicx,color}
\usepackage{amssymb,amsmath,bm}
\usepackage{textcomp}
\usepackage{array}
\usepackage{stfloats}
\usepackage{epsfig,epstopdf}

\begin{document}

\title{Decision Making Based on Cohort Scores for Speaker Verification}

\author{%
\authorblockN{%
Lantian Li\authorrefmark{2}, Renyu Wang\authorrefmark{3}, Gang Wang\authorrefmark{2}, Caixia Wang\authorrefmark{2}, Thomas Fang Zheng$^*$\authorrefmark{2}
}

\authorblockA{%
\authorrefmark{2}
Center for Speech and Language Technologies, Division of Technical Innovation and Development, \\
Tsinghua National Laboratory for Information Science and Technology;\\
Center for Speech and Language Technologies, Research Institute of Information Technology;\\
Department of Computer Science and Technology, Tsinghua University, Beijing, China\\
}

\authorblockA{%
\authorrefmark{3}
School of Linguistic Sciences, Jiangsu Normal University, China}

\authorblockA{%
\authorrefmark{1}
Corresponding Author E-mail: fzheng@tsinghua.edu.cn} \\

}
\maketitle

\thispagestyle{empty}

\begin{abstract}

Decision making is an important component in a speaker verification system. For the conventional GMM-UBM
architecture, the decision is usually conducted based on the log likelihood ratio of the test utterance against
the GMM of the claimed speaker and the UBM. This single-score decision is simple but tends to
be sensitive to the complex variations in speech signals (e.g. text content,
channel, speaking style, etc.). In this paper, we propose a decision making approach based on
multiple scores derived from a set of cohort GMMs (cohort scores). Importantly,
these cohort scores are not simply averaged as in conventional cohort methods; instead,
we employ a powerful discriminative model as the decision maker.
Experimental results show that the proposed method delivers substantial performance improvement
over the baseline system, especially when a deep neural network (DNN) is used as the
decision maker, and the DNN input involves some statistical features derived from the cohort scores.

\end{abstract}

\section{Introduction}

Speaker verification aims to verify claimed identities of speakers, and has gained great popularity in a wide range of applications including access control, forensic evidence provision and user authentication.
After decades of research, lots of popular speaker verification approaches have been proposed, such as Gaussian mixture model-universal background model (GMM-UBM)~\cite{sid:reynolds00},
joint factor analysis (JFA)~\cite{kenny2007joint} and its `simplified' version, the i-vector model~\cite{dehak2011front}.
Accompanied with these models, various back-end techniques have also been proposed to promote the discriminative capability
for speakers, such as within-class covariance normalization (WCCN)~\cite{hatch2006within}, nuisance attribute projection (NAP)~\cite{campbell2006support} and probabilistic LDA (PLDA)~\cite{prince2007probabilistic}, etc.
These methods have been demonstrated to be highly successful.
Recently, deep learning has been applied to speaker verification and gained much interest~\cite{kennydeep,Ehsan14}.

Within a speaker verification system, decision making is an important component~\cite{campbell1997speaker}.
To make a decision, the verification system first determines a score for the test utterance that reflects
the confidence that the utterance is from the claimed speaker, and then compares the score with a predefined threshold.
In a typical GMM-UBM system, the score is often computed as the log likelihood ratio that the test utterance
being generated from the GMM of the claimed speaker and the UBM.
This single-score decision is simple and efficient, but it tends to be quite sensitive to variations in
speech signals, e.g., in terms of text contents, channel conditions and speaking styles. This sensitivity
means that choosing an appropriate threshold is rather difficult, or leading to error-pron decisions.

To deal with this score variation, various score normalization techniques have been proposed.
Most of the normalization approaches, according to~\cite{sid:auckenthaler00}, can be
explained using the Bayes' theorem. Among these approaches the cohort normalization
is particular interesting. This approach chooses a set of cohort speakers who are close to the genuine speaker,
and for each test utterance, it computes a set of `cohort scores' on the models of these speakers.
These cohort scores then replace the UBM to normalize the score of the test utterance against
the claimed speaker~\cite{finan1992,rosenberg1992}. Using cohort models tends to model the
alternative hypothesis more accurately, due to its more flexible structure compared to a single UBM.
However, the existing methods based on cohort models do not fully utilize the information
involved in the cohort scores: they are just simply averaged to normalize the target score, which
is still a single-score approach.

This paper presents a new cohort approach that utilizes the cohort scores in a more effective way.
Specifically, we propose to make decisions on the whole cohort scores (formulated as a score vector),
and employ a powerful discriminative model to make the decision.
Our assumption is that the knowledge involved in the cohort scores is more than a mean average,
but as complex as their distributions, their ranks, spanning areas, etc. Fully utilization of
these rich information results in a true multi-score decision making, which is expected to be
more reliable than the traditional single-score approach.

The technique presented in this paper involves three steps:
(1) Firstly, a set of cohort models is constructed by a clustering algorithm;
(2) Secondly, for each test utterance, scores are estimated among the claimed speaker GMM, the global UBM and
the cohort GMMs;
(3) Finally, a classification model (SVM or DNNs) is employed to make the decision based on some features
derived from the scores derived above.


The layout of this paper is organized as follows. Section~\ref{sec:framework} presents the proposed cohort-based
decision making framework. The experiments are presented in Section~\ref{sec:exp}, and Section~\ref{sec:conl}
concludes the paper.

\section{Cohort-based decision making framework}
\label{sec:framework}

\begin{figure}[htp]
\centering
\includegraphics[width=0.85\linewidth]{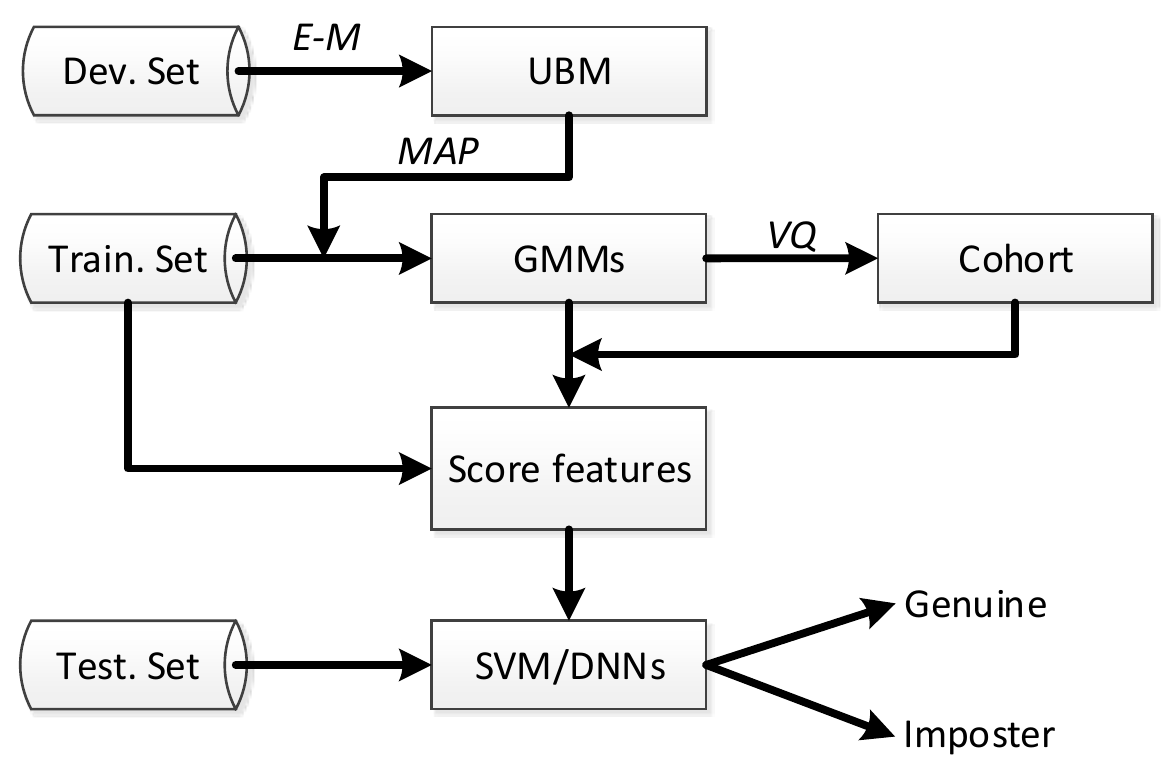}
\vspace{-2mm}
\caption{The cohort-based decision making framework.}
\label{fig:framework}
\end{figure}

In a typical GMM-UBM speaker verification system, the score likelihood ratio of a test utterance is computed over the GMM of the claimed speaker model and UBM.
Then the likelihood ratio will be compared with a predefined threshold.
If it is higher than the threshold, the test utterance will be accepted, else rejected.
We argue that this naive decision making approach is unreliable and less robustness because this likelihood ratio only describes the distance between the claimed GMM and UBM,
and it does not make use of the world speakers and corresponding score information.
Therefore, we design a cohort-based decision making framework, as shown in Fig.~\ref{fig:framework}.
This framework is made up of three parts: cohort selection, feature design and discriminative model training.

\subsection{Cohort selection}

A vector quantization (VQ) method ~\cite{gray1984} based on the K-means algorithm was utilized to conduct the speaker model clustering.
The centroid of each cluster represents a reference speaker, and all the reference speakers build the `cohort'.
We chose a weighted K-L distance to measure the distances among Gaussian mixture models, given by:

\begin{equation}
   D(\lambda_{1},\lambda_{2}) = \sum_{i=1}^{M} w_{i} KL(N_{i}^{1}, N_{i}^{2})
   \label{eq:gaussian}
\end{equation}

\begin{equation}
   KL(N_{1}, N_{2}) = \frac{1}{2} (\mu_{1} - \mu_{2})^{T}(\Sigma_{1}^{-1} - \Sigma_{2}^{-1})(\mu_{1} - \mu_{2})
   \label{eq:kl}
\end{equation}

\noindent where $\lambda_{1}$ and $\lambda_{2}$ are two Gaussian mixture models, and $w_{i}$ is the weight of $i^{th}$ Gaussian component.
Note that, for fast computation, only the mean parameters are adapted in the GMM-MAP process, while the weights and variances of the GMMs are the same as UBM.
Equ.~\ref{eq:kl} is used to measure the distance between two multi-dimensional Gaussian distributions.

Given a set of speaker GMMs $\lambda$ = ($\lambda_{1}, \lambda_{2}, ..., \lambda_{n}$) and $\lambda_{c^{i}}$ that is the cluster centroid where speaker $i$ is assigned to.
The optimization objective is to minimize the within-class cost $J$,
and finally each cluster centroid is regarded as one `cohort' model.

\begin{equation}
   J = \frac{1}{N} \sum_{i=1}^{N} D(\lambda_{i}, \lambda_{c^{i}})
   \label{eq:cost}
\end{equation}

\subsection{Feature design}

Once the cohort models (CGMMs) have been determined, a set of cohort scores are calculated on the claimed speaker GMM, UBM and CGMMs respectively for each test utterance.
We seek to use these cohort scores to explore some potential knowledge and design more discriminative features on genuine and imposter speaker models.
In this part, three cohort-based score features are discussed.

\subsubsection{Cohort-based score normalization}
The inspiration of this feature comes from the conventional score normalization techniques ~\cite{sid:auckenthaler00}.
For a test feature vector $X$, the normalized score $\widetilde{L}_\lambda(X)$ is given as follows:

\begin{equation}
   \widetilde{L}_\lambda(X) = \frac{L_\lambda(X) - \mu_\lambda}{\sigma_\lambda}
   \label{eq:norm}
\end{equation}

\noindent where $\lambda$ represents a claimed speaker model, and $\mu_\lambda$, $\sigma_\lambda$ is estimated from the cohort scores.

\subsubsection{Rank position}

Assuming the size of cohort is $K$, for each test trial, a ($1$+$K$)-dimensional score vector is calculated based on GMM and CGMMs.
And we think that the likelihood scores on the genuine speaker GMMs are at the top-rank position in the ($1$+$K$)-dimensional score vector,
while for the imposter speakers, it lies in a random rank position.

\subsubsection{Rank of score differences}
\label{sec:rankdiff}

Similar assumption with the rank position, we also believe that the distribution of cohort scores on the genuine speaker models is different from that on imposter speaker models.
For each test utterance, the score feature is calculated by subtracting the likelihood score on the claimed speaker GMM from that on each cohort CGMM.
It describes a high-dimensional cohort-based score distribution instead of the UBM space.
After ranking it, this score feature also covers the information of rank position, and has strong discriminability on genuine and imposter speaker models.
This assumption will be verified in Section~\ref{sec:feature}.


\subsection{Discriminative model training}

Based on these features derived from the cohort scores, discriminative models (e.g., support vector machine (SVM) and deep neural networks (DNNs) can be directly optimize
with respect to the speaker verification task, i.e., the genuine/imposter speaker decision.

\section{Experiments}
\label{sec:exp}

\subsection{Database}

The experiments are performed on a database called `\emph{CSLT-DSDB}' (Digit String Database) that
was jointly created by CSLT (Center for Speech and Language Technologies), Tsinghua University and
Beijing d-Ear Technologies, Co. Ltd. The text of all recordings is the text-prompted digit strings.
The recordings were conducted using different mobile microphones, sampled at $16$ kHz with $16$-bit precision.

\begin{itemize}
   \item Training set: It contains an approximate size of $1$ GB data (about $200$ males and $200$ females)
recorded in an ordinary office environment. And it is used for the UBM training.
\end{itemize}

\begin{itemize}
    \item Development set: It contains $280$ enrollment utterances covering $145$ speakers and $2,874$ test utterances. And it is used for cohort selection and feature design.
\end{itemize}

\begin{itemize}
    \item Evaluation set: It involves $92$ speakers. For each speaker, there are text-prompted digit strings of about $40$ seconds in length for speaker model training;
and $8$-$16$ randomly generated digit strings each of which is an 8-digit string for verification.
There are overall $1,220$ test utterances and $1,220$ target trials and $111,020$ non-target trials.
\end{itemize}

\subsection{Experimental setup}

The acoustic feature was the conventional $39$-dimensional Mel frequency cepstral coefficients (MFCC), which involves $13$-dimensional static components plus the first and second order derivatives.
The UBM consisted of $256$ Gaussian components and was trained with the training set. Note that this setting is `almost' optimal in our experiments,
i.e., using more Gaussian components cannot improve system performance.
And the baseline of GMM-UBM system on the evaluation set was $1.621\%$ in terms of EER (Equal Error Rate).

\begin{figure}[htp]
\centering
\includegraphics[width=0.85\linewidth]{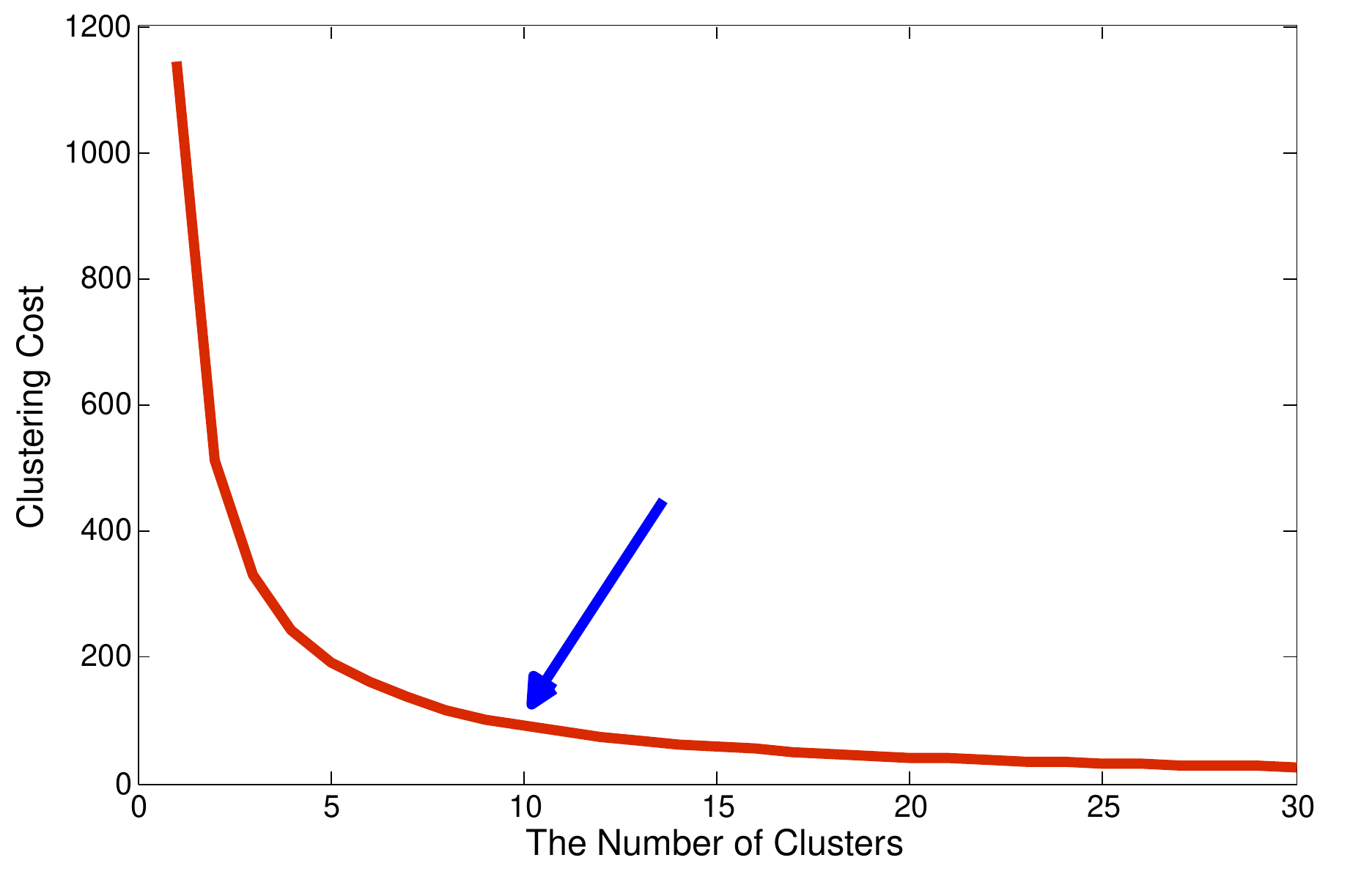}
\vspace{-2mm}
\caption{The function between the number of clusters and the clustering cost.}
\label{fig:cost}
\end{figure}

Besides, with the maximum a posterior (MAP) algorithm, $280$ speaker GMMs were adapted from UBM. And The K-means algorithm was used to cluster the $280$ speaker GMMs into a suitable cohort.
Fig.~\ref{fig:cost} presents the function between the number of clusters and the clustering cost $J$.
It can be observed that when the number of clusters exceeds $10$, the clustering cost $J$ has already been converged. Therefore, the size of cohort was set to $10$.

In order to select the discriminative score feature, $6,115$ target trials and $5,748$ imposter trials were selected from the development set.
Considering the unbalanced data problem \footnote{The number of target samples and imposter samples will be highly unbalanced,
one or some few target samples against large amount of imposter samples. And learning from such unbalanced data will result in biased SVM/DNNs models.},
only the top two scores were selected from all the imposter speaker models.

\subsection{Feature design}
\label{sec:feature}

\subsubsection{Cohort-based score normalization}

According to Equ.~\ref{eq:norm}, the normalized score for each test was calculated, and the system performance was $1.639\%$ in EER.
It shows reasonable performance and can be considered as an available score feature.

\subsubsection{Rank position}

\begin{figure}[htp]
\centering
\includegraphics[width=0.85\linewidth]{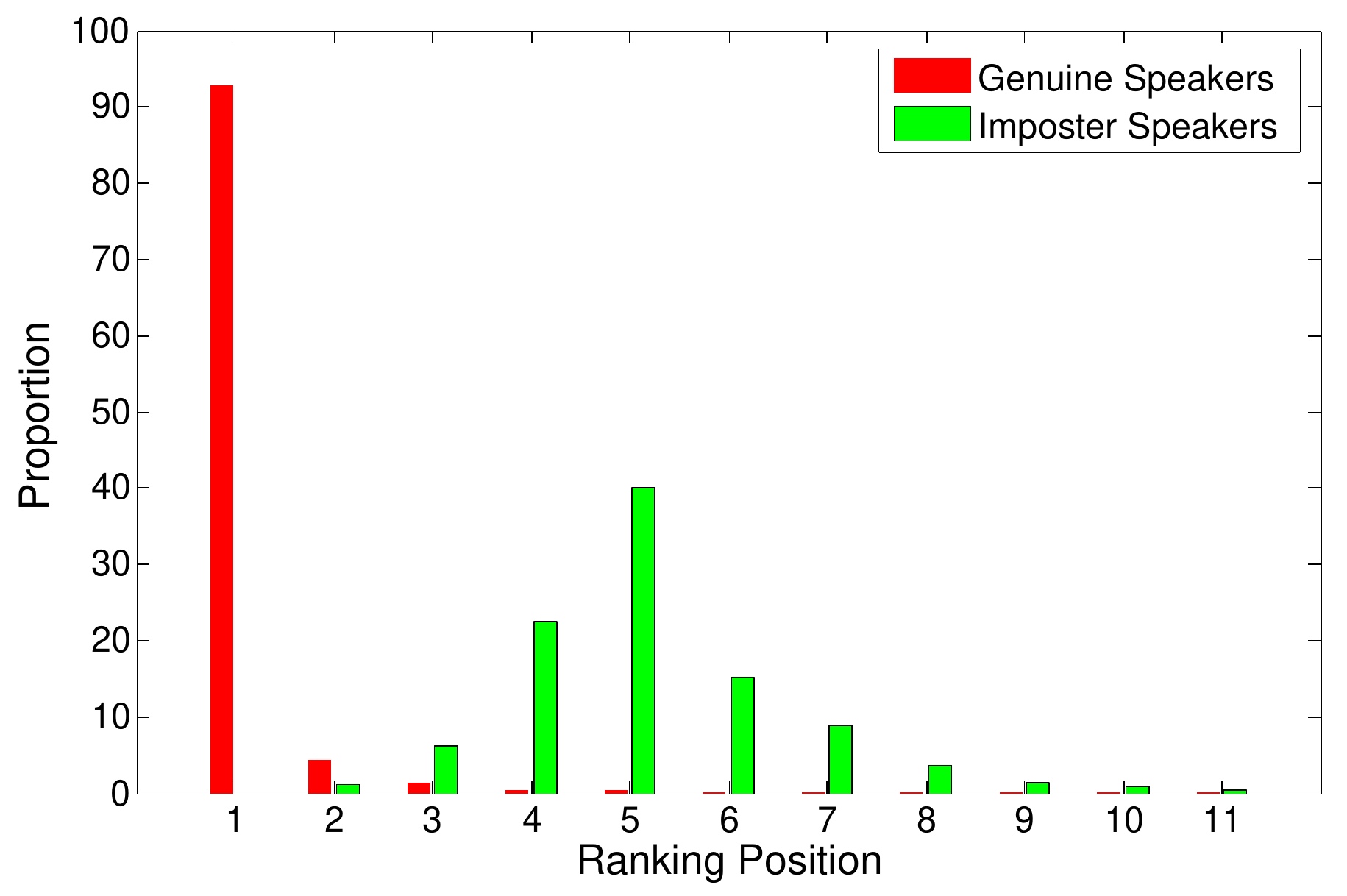}
\vspace{-2mm}
\caption{The statistical histogram of rank-position distribution.}
\label{fig:rank}
\end{figure}

From Fig.~\ref{fig:rank}, we observed that this rank position has certain discriminability.
Nearly all the likelihood scores on the genuine speaker GMMs are at the first rank position,
while for imposter speaker models, the rank position distribution approximately satisfies a Gaussian distribution with the mean of $5$.

\subsubsection{Rank of score differences}

\begin{figure}[htp]
\centering
\includegraphics[width=0.85\linewidth]{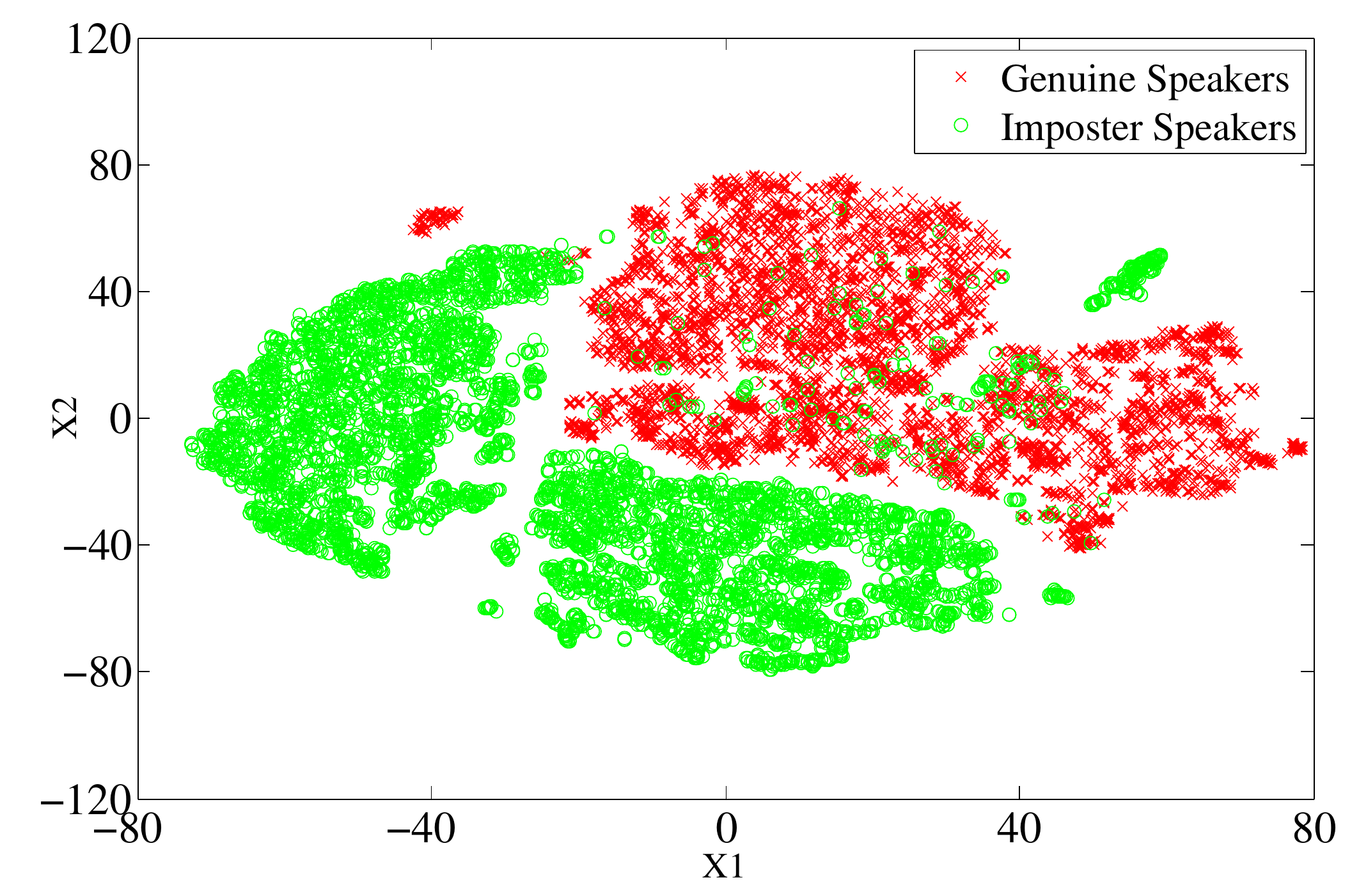}
\vspace{-2mm}
\caption{The distribution of the `rank of score differences' on genuine speakers and imposter speakers.}
\label{fig:sdf}
\end{figure}

To provide an intuitive understanding of the discriminative capability of this feature, the rank of score differences of all the test trials are plotted in a two-dimensional space via T-SNE ~\cite{saaten2008}.
As shown in Fig.~\ref{fig:sdf}.

It can be seen that there exists a distinct non-linear boundary between genuine speaker models and imposter speaker models.
That is to say, this `rank of score differences' has strong discriminative capability on genuine and imposter speaker models.

\subsection{Discriminative model training}

With these cohort-based score features, discriminative models can be optimized with respect to discriminating the genuine/imposter speakers.
In this paper, both the SVM and DNNs models were trained as the decision maker for speaker verification system.

\subsubsection{SVM-based scoring}

The SVMs were trained for each cohort-based score feature with the linear kernel function.
Results are shown in Table~\ref{tab:svm} on condition of C1-C3. Note that `norm' is the `Cohort-based score normalization', `r-pos' is the `Rank position', `r-diff' is the `Rank of score differences'
and $\surd$ represents that related features are chosen as the input of SVMs.

\begin{table}[htb]
\begin{center}
\caption{The SVM-based discriminative scoring evaluation system.}
          \begin{tabular}{l|c|c|c|c|c}
           \hline
                 Condition   &   score    &   norm      &   r-pos     &     r-diff      &   EER(\%)    \\
           \hline
                   C1        &  $\surd$   &   $\surd$   &    --       &      --         &  1.598       \\
           \hline
                   C2        &  $\surd$   &   --        &   $\surd$   &      --         &  1.574       \\
           \hline
                   C3        &  $\surd$   &   --        &   --        &     $\surd$     &  1.475       \\
            \hline
            \hline
                   C4        &  $\surd$   &   $\surd$   &   $\surd$   &      --         &  1.625       \\
            \hline
                   C5        &  $\surd$   &   $\surd$   &   --        &      $\surd$    &  1.475       \\
            \hline
                   C6        &  $\surd$   &   --        &   $\surd$   &      $\surd$    &  1.475       \\
            \hline
                   C7        &  $\surd$   &   $\surd$   &   $\surd$   &      $\surd$    &  1.479       \\
            \hline
          \end{tabular}

          \label{tab:svm}
\end{center}
\end{table}

\subsubsection{DNN-based scoring}

The DNN models were trained with these cohort-based score features, and the decision was made by logistic regression model at the soft-max layer.
Note that for different input feature, the experimental results can be optimized with tuning of the DNNs structure such as the number of hidden units and hidden layers.
Whereas, in order to unify the experimental configuration, we just set the number of hidden layer units $10$ times as much as the dimension of input features, and there is only $1$ hidden layers.
The results are shown in Table~\ref{tab:dnn} on the condition of C1-C3.

\begin{table}[htp]
\begin{center}
\caption{The DNN-based discriminative scoring evaluation system.}
          \begin{tabular}{l|c|c|c|c|c}
           \hline
                 Condition   &   score    &   norm      &   r-pos     &     r-diff      &   EER(\%)    \\
           \hline
                   C1        &  $\surd$   &   $\surd$   &    --       &      --         &  1.556       \\
           \hline
                   C2        &  $\surd$   &   --        &   $\surd$   &      --         &  1.639       \\
           \hline
                   C3        &  $\surd$   &   --        &   --        &     $\surd$     &  1.148       \\
            \hline
            \hline
                   C4        &  $\surd$   &   $\surd$   &   $\surd$   &      --         &  1.639       \\
            \hline
                   C5        &  $\surd$   &   $\surd$   &   --        &      $\surd$    &  1.230       \\
            \hline
                   C6        &  $\surd$   &   --        &   $\surd$   &      $\surd$    &  2.049       \\
            \hline
                   C7        &  $\surd$   &   $\surd$   &   $\surd$   &      $\surd$    &  1.077       \\
            \hline
          \end{tabular}
          \label{tab:dnn}
\end{center}
\end{table}

\subsubsection{Feature combination}

From Table~\ref{tab:svm} and Table~\ref{tab:dnn}, it can be seen that in condition C1-C3, both the SVM- and DNN-based scoring offer clear performance
improvement than the GMM-UBM baseline $1.621\%$. Therefore, a feature combination scheme was proposed by concatenation these score features together. Experiment results are shown on condition of C4-C7.
It can be observed that the performance of this simple feature combination is inconsistent, and we attribute it to the feature redundancy because all these features are embedded from the cohort scores.
Besides, the overall feature combination C7 on DNN-based scoring system obtains the best performance.

\section{Conclusions}
\label{sec:conl}

This paper presents a decision making method based on cohort scores instead of the traditional single decision score.
Some potential discriminative features are embedded from cohort scores, and then more powerful discriminative models are trained as the decision maker.
Experimental results show that the proposed `rank of score differences' with SVM/DNN-based scoring model can obtain stable and better system performance than the GMM-UBM baseline.
Moreover, a feature combination scheme is proposed to further improve system performance.
Future work involves designing more robustness score-level discriminative features and more reasonable cohort selection approaches.

\section*{Acknowledgment}

This work is supported by the National Natural Science Foundation of China under Grant No. 61371136 and No. 61271389, it was also supported by the National Basic Research Program  (973 Program) of China under Grant No. 2013CB329302.

\bibliographystyle{IEEEtran}
\bibliography{cohort}

\end{document}